\documentstyle[prl,aps]{revtex}
\begin{document}

\title{Three Quantum Aspects of Gravity}

\author{D. V. Ahluwalia}

\address{Mail Stop H--846, Physics (P--25)  Division,
Los Alamos National Laboratory, Los Alamos, NM 87545, USA\\
and, Global Power Division, ANSER, S-800, 1215 Jefferson Davis Highway,
Arlington, VA 22202, USA
}
\maketitle
\vskip 0.25cm

\begin{abstract}
It is argued that (a) In the quantum realm test--particle masses
have non--trivial observability which induces a non--geometric element in
gravity , (b) Any theory of quantum gravity, on fundamental grounds, 
must contain an element of non--locality that makes position measurements 
non--commutative, and (c) The classical notion of free fall does not 
readily generalize to the quantum regime.
\end{abstract}

\vskip 0.5cm
I came across Professor Ta-You Wu
in 1976 at the SUNY at Buffalo. He was not only an
energetic man in the physics corridors of Fronczak Hall,
but the most energetic one.
I never saw him taking the elevators, he always walked, no! he ran the 
steps --- he seemed to carry in his mind restless thoughts 
and had the apparent need to immediately share them,
in their raw and unedited intensity,  with
either a colleague or a student. So these thoughts became the fuel,
or so seemed to me, that they fueled his steps.
His charm, his child-like affection for physics,
his letters written to me while flying at some high altitude in a plane 
and talking about possible unification of gravity and electromagnetism, 
are still very fresh in my memory.\footnotemark[1]
\footnotetext[1]
{I very much regret that these Wu letters, full of ideas and love for
physics, have gotten
lost as I moved from one place to another in my own learnings of
physics.}
So much that despite the fact that
he and I have had no occasion to meet in person again
since those Buffalo days I recently dreamed 
of him. It was out of such a night 
dream that it occurred to me, at the invitation
of Jen--Chieh Peng, that I write him a letter. What follows is a
letter, written sitting at a lower altitude of 7500 feet in Los
Alamos --- a letter that I hope will invite comments and wisdom of
my teacher under whom I began to learn physics 
and with whom
I published my first research paper \cite{first}.\footnotemark[2]
\footnotetext[2]{As Professor Wu left Buffalo, I took a detour to 
film school, only returning to physics to obtain my Ph.D. in 1991.}

For historical reasons the Einstein's theory of gravitation
remains clouded
in mystery for most students of physics. This  myth and 
mystery, combined with the theory's mathematical beauty, 
has made, at times,  its experts arrogant and rigid
in their purely geometrical thinking. Yet, there remain serious
physicists who warn against too geometrical a view \cite{SW,RF}.
Steven Weinberg  \cite[p. vii]{SW}, for instance, notes that
`` ... too great an emphasis on geometry can only obscure the deep
connection between gravitation and the rest of physics.''
Moreover, the mere fact that gravitational interaction between 
two electrons is about forty orders of magnitude weaker than
the electromagnetic intercation does not imply, contrary to
the claims by even so distinguished a physicist as Feynman \cite{RF},
that quantum effects of gravity cannot be studied in the
Earth based laboratories. The gravity of Earth is strong enough
to result in large observable gravitationally induced quantum
interference phases \cite{COW,atomic}.

The geometrical interpretation of gravity in the classical
realm, for instance, arises from the fact that  all
clocks red shift identically when introduced in a given gravitational
environment. 
Here,
I review simple conceptual
considerations and show that already known elements of quantum
mechanics and gravity require that not all clocks red shift
identically in the gravitational environment of a rotating source.
This introduces an element of non--universality in the red shifting of
the clocks and thus  
 suggests a non--geometric element in gravity
in the quantum context. 
Second, I point out that the ``collapse of wave function''
and the associated  gravitational considerations, imply an intrinsic
element of nonlocality in gravity in the quantum regime.
These two results place severe constraints on how we imagine the
ultimate unification of gravity with other interactions of nature.
Finally, I examine in some detail the extension of the 
notion of free fall to the quantum realm.

Following Sakurai
\cite{JJS} I treat gravity on equal footing with other interactions. 
This is justified, at the very minimum, in the weak--field limit.
The weak--field limit is sufficient, and indeed desirable,
\footnotemark[3]
\footnotetext[3]{The desirability, in part, arises from the fact that a 
purely geometric framework can hardly be expected to hint at a 
non-geometric element. The ultimate relevance of the arguments that
are presented here, therefore, lies in their suggestion of
experiments. Such experiments are already underway \cite{COW,atomic}.}
for the conceptual matters that I wish to study. 
In this framework, the classical motion of a test particle of 
mass $m$ in the  gravitational field produced by a 
non--rotating source of mass $M$   
is governed by
\begin{equation}
m\frac{\mbox{d}^2 \bbox{x}}
{\mbox{d} t^2} = -\,m\bbox{\nabla} \phi_{grav}\quad, \label{cl}
\end{equation}
while in the non-relativistic quantum regime, one finds
\begin{equation}
\left[ -\left(\frac{\hbar^2}{2 m}\right)
\bbox{\nabla}^2 + m \phi_{grav}\right]
\psi = i\hbar \frac{\partial\psi}{\partial t}\quad .\label{qu}
\end{equation}

The empirically observed equality of the inertial and gravitational
masses enters explicitly both in  Eqs.~(\ref{cl}) and (\ref{qu}), however, 
this happens in two different ways:
\begin{enumerate}
\item
The inertial mass appears on the L.H.S. of Eq.~(\ref{cl}) as well as
in the kinetic term of Eq.~(\ref{qu}), while the 
gravitational mass enters the R.H.S. of Eq.~(\ref{cl}) as a force 
on the one hand,
and as an   interaction--energy term in Eq.~(\ref{qu}), on the other hand.

\item
The gravitational potential $\phi_{grav} = -GM/r $ 
emerges within the weak--field limit of Einstein's theory of gravitation
and is deduced directly from the equivalence principle. 
\end{enumerate}

Sakurai noted that while the test--particle mass cancels out on
both sides of the classical equation of motion Eq. (\ref{cl}), this
is not the case for the quantum mechanical Shroedinger equation. 
As a consequence, the experimentally observed gravitationally induced quantum 
interference in the  Collela, Overhauser, and Werner (COW) experiment on
neutron interferometry carries explicit information on the
test--particle mass (i.e., neutron mass) \cite{COW}.

In addition, I note that 
it is $\bbox{\nabla} \phi_{grav}$ that governs the physics
of  
Eq. (\ref{cl}) while in Eq. (\ref{qu}) it is $\phi_{grav}$ that
is directly operative. 
However, even this
necessary difference is not always
sufficient  for physically
observable consequences.
For gravitational  environment that is characterized by 
an essentially constant $\phi_{grav}$ the classical effects vanish
while quantum effects under certain circumstances do not (see below).

Following  these observations and inspired by the 
COW experiment, I recently suggested in collaboration with Burgard
that if one considers a state that is a linear superposition of mass 
eigenstates \footnotemark[4]
\footnotetext[4]{
Such states indeed exist in Nature:  the $K$--$\overline{K}$ system, and
neutrinos that are now experimentally indicated to be
linear superposition of mass eigenstates \cite{LSND,Kamioka}, provide
two such examples.}
then each of the 
mass eigenstate picks up a mass--dependent  gravitationally induced phase 
\cite{grf96,ov}.
This phase, which would have been
a global factor, and hence 
devoid of observability for an isolated mass eigenstate, 
now gives 
rise to observable relative phases between the various mass eigenstates.
Specifically, for neutrinos I noted that the phenomena of neutrino oscillations
provides a flavor--oscillation clock and the 
mass--dependent  gravitationally induced phases make this clock red shift as 
expected on the basis of Einstein's theory of gravitation.

However this universality of the red shift, which is
so important for the geometrical interpretation of gravity in the classical
realm, breaks down if one considers a linear superposition of spin 
projections (with different or equal masses) in the vicinity of  a rotating
gravitational source. 
The reason for this break down of the 
universality lies in the spin--projection dependence of the gravitationally
induced quantum phases.
The details of this argument I presented
in Ref. \cite{grf97}.

Sakurai had already noted in Ref. \cite[p. 126]{JJS} that 
``because mass does not appear in the equation of a particle trajectory, 
gravity in classical mechanics is often said to be a purely geometric theory.''
Now my recent work summarized here shows that because
trajectory of quantum test particles can carry flavor--oscillation
clocks (whose beating depends on masses and spin projections
 of the superimposed mass eigenstates) the non--universal red shifting 
of these clocks explicitly depends on the test particle. Hence,
I suggest that in the quantum realm  the theory of gravity
contains a non-geometrical element.

The second observation that I wish to report here is that
the collapse of a wave function is associated with the 
collapse of the energy--momentum tensor. Since it is the 
energy--momentum tensor that determines the spacetime metric, 
the position measurements alter the spacetime metric in  a fundamental and
unavoidable manner. Therefore, 
in the absence of external gravitating sources (which otherwise
dominate the spacetime metric), it matters, in principle,
in what order we make position measurements of particles \cite{grf94}. 
Quantum mechanics
and gravity intermingle in such a manner as to
 make position measurements non--commutative.
This then brings to our attention another intrinsic element of gravity in the
quantum realm, the element of non--locality.

Whether this non-locality results in the violation of the CPT symmetry
is not yet known \cite{CPTa,CPTb,CPTc}.

As a final observation,
I note that the 
classical notion of  the free fall in essence is the local equivalence
at a spacetime point of acceleration $\mbox{d}^2\bbox{x}/\mbox{d}t^2$ and  
$\bbox{g}\equiv - \bbox{\nabla} \phi_{grav}$. This equivalence
remains invariant under
\begin{equation}
\phi_{grav}\,\rightarrow\,\phi_{grav}+\phi^0_{grav}
\quad,\label{tran}
\end{equation}
 where $\phi^0_{grav}$
is essentially constant over the 
spacetime region of experimental interest.
However, the gravitationally induced quantum mechanical 
relative phases
are not invariant under such a transformation, and the readings
of clocks in classical free
fall may not be considered to give readings that correspond to
$\phi_{grav} = 0$. Clocks in free
fall give   readings that correspond to an absence of
gravitationally induced accelerations (not phases), i.e.,
to $\bbox{\nabla} \phi_{grav}=\bbox{0}$. 
Mathematically, (a) The transformation (\ref{tran}) does not
alter Eq. (\ref{cl}) whereas it changes Eq. (\ref{qu}), and (b)
Vanishing of a gravitationally induced acceleration, i.e.,
$\bbox{g}=\bbox{0}$, in  a given frame,
 does not imply vanishing of the gravitationally
induced phases (in the same frame).
It is not clear
what is the appropriate generalization of the classical free fall
to the quantum realm. 
Many quantum mechanical clocks are driven  
by mass, or energy, dependent relative phases and red shift via
gravitationally induced
phases that depend on $\phi$.  The 
red shift of clocks based on quantum mechanical phases can be
measured in systematic terrestrial experiments.  
Classical clocks, on other hand,  are insensitive in a free fall
to the existence of an essentially constant $\phi^0_{grav}$ --- so, at least,
is the case within the general relativistic framework.
\footnotemark[5]
\footnotetext[5]
{In the free fall the general--relativistic 
spacetime metric is $\eta_{\mu\nu}$,
in the notation of Ref. \cite{SW}, 
and therefore $\phi^0_{grav}$ does not contribute
to red shift of the clocks within general relativistic framework.} 
The question raised here is far from being irrelevant experimentally.
In the 
latest neutron interferometry experiments a discrepancy between theory and
experiments continues to exist at the several standard deviation level 
\cite{COWd}. I suggest that the observed discrepancy may be
pointing not towards some yet unknown systematic errors in the
experiments but  is indicative of 
  a non--vanishing $\phi^0_{grav}$. A non--zero
$\phi^0_{grav}$, that is essentially constant over the
scale of the solar system (say), can easily  arise from the cosmological
distribution of matter.

I conclude by noting that there is a  non--geometric element in
gravity in the quantum realm and that any theory of quantum gravity, 
on fundamental grounds, must contain an element of non--locality that
makes position measurements non--commutative.
The fundamental notion of free fall, so intricately connected
to the principle of equivalence, itself requires further study
in the quantum regime. The general observation can therefore be made that the
general theory of relativity arose out of the data  on planetary
orbits. These orbits are determined entirely by the gradient
of the gravitational potential, and are insensitive to any contribution
to the potential by the cosmological sources. This insensitivity 
arises because on the length scales of the solar system, or even galactic
regions of spacetime, the cosmological
contribution to the gravitational potential is constant to a very
good accuracy. However, quantum effects remain capable of measuring
this cosmological potential. But, general theory of relativity rules out
any physical consequences from an essentially constant gravitational
potential. Therefore, the inevitable conclusion is reached that in the
quantum realm general relativity must suffer fundamental changes.
The three quantum aspects of gravity discussed here point towards the
conceptual nature of these changes.

{\em This work was done, in part, under the auspices of the 
U.S. Department of Energy.}

\end{document}